\documentclass[runningheads]{llncs}
\usepackage{graphicx}
\usepackage{xcolor}
\usepackage{booktabs}
\usepackage{array}
\usepackage{colortbl}
\usepackage{filecontents}
\usepackage{longtable}

\newcommand{\levelone}[1]{\textbf{#1}}
\newcommand{\leveltwo}[1]{\textit{\hspace{2mm}-- #1}}

\begin{document}

\title{A Comprehensive View on Quality Characteristics of the IoT Solutions}
%\author{Author A\inst{1,2}\orcidID{0000-0000-0000-0000} \and
%Author B\inst{2}\orcidID{0000-0000-0000-0000} \and
%Author C\inst{1} \orcidID{0000-0000-0000-0000} \and 
%Author D\inst{1}\orcidID{0000-0000-0000-0000}}

\author{Miroslav Bures\inst{1,2}\orcidID{0000-0002-2994-7826} \and
Xavier Bellekens\inst{2}\orcidID{0000-0003-1849-5788} \and
Karel Frajtak\inst{1} \orcidID{0000-0003-4133-2805} \and 
Bestoun S. Ahmed\inst{1}\orcidID{0000-0001-9051-7609}}

\authorrunning{Miroslav Bures et al.}
%\institute{University A\and
%University B}

\institute{FEE, Czech Technical University in Prague, Czech Republic \email{\{miroslav.bures,frajtak,albeybes\}@fel.cvut.cz}\and
School of Design and Informatics, Abertay University, UK
\email{x.bellekens@abertay.ac.uk}}

\maketitle              % typeset the header of the contribution
\begin{abstract}
Categorization of quality characteristics helps in a more effective structuring of the testing process and in the determination of properties, which can be verified in the system under test. In the emerging area of Internet of Things (IoT) systems, several individual attempts have been made to summarize these aspects, but the previous work is rather heterogenic and focuses on specific subareas. Hence, we consolidated the quality characteristics into one unified view, which specifically emphasizes the aspects of security, privacy, reliability and usability, as these aspects are of often quoted as major challenges in the quality of contemporary IoT systems. The consolidated view also covers other areas of system quality, which are relevant for IoT system testing and quality assurance. In the paper, we also discuss relevant synonyms of particular quality characteristics as presented in the literature or being used in the current industry praxis. The consolidated view uses two levels of characteristics to maintain a suitable level of granularity and specificity in the discussed quality characteristics.

\end{abstract}
\section{Introduction}

To measure the quality of a System Under Test (SUT), various quality characteristics are being employed as standard industry practice, for instance, \cite{jung2004measuring,van2013tmap,guceglioglu2005using}. These characteristics are covering various quality aspects of the SUT, spanning from the presence of defects in SUT functionality to broader issues, ranging from usability and maintainability to the testability of the systems \cite{van2013tmap}.

The importance of these characteristics lays in several functions: 

%% I will re-modify this when my internet connection comes back up. Xavier
\begin{enumerate}
\item They serve as managerial tools to measure the quality of SUT and contribute to making the quality assurance process more effective;
\item They emphasize different quality aspects besides the simple presence of software defects. Hence, they lead system engineers to focus on aspects like testability, maintainability, scalability or another, which are not directly quantified by the presence of defects explicitly reported by the testing teams. However, these can have a  significant impact on project or product success or failure;
\item They help setting-up an efficient test strategy for particular SUT, but most importantly they help managerial decisions regarding the quality aspects that are important and inform on which techniques shall be employed to prove SUT quality;
\item Considering the fact that the security and privacy are discussed as the main challenges of the current IoT solutions \cite{kiruthika2015software,bures2018}, proper quality characteristics may help reducing cyber--security and privacy risk by revealing flaws and reducing the attack surface by ensuring that the correct operations are executed. 
\end{enumerate}

In the software testing and quality assurance industry, several sets of quality characteristics have been established and used. As an example, we can give \textit{ISO/IEC 9126}, later replaced by \textit{ISO/IEC 25010:2011} \cite{jung2004measuring} or \textit{TMap Next} \cite{van2013tmap}. As IoT systems differ from web-based software enterprise systems in a number of points it also brings several challenges specific to IoT infrastructures \cite{kiruthika2015software,bures2018}, it is, therefore, relevant to revise these quality characteristics and quality metrics for IoT systems and to capture their specificities. A first attempts has already been made \cite{zheng2014,li2012,TMAP,islam2011,Ikeda2016,Banerjee2017,manuel2015,han2014,buyya2009,gomes2018,abbadi2011security,badenhorst2013developing,sollie2005security}, however these classifications focus specifically on heterogenic areas of IoT systems, applications and viewpoint on the system quality. Hence, a more consolidated system is required,  discussing the particulars of IoT domains and their intrinsic nature. 

Generally, we need to distinguish between quality characteristics and quality metrics. \textbf{Quality characteristic} is a general category, describing a particular viewpoint on the SUT, which can be used in the test planning, test strategy or test reporting. It is not defined by a particular formula which employs particular elements and quantities form the SUT model or facts from the test management process. Differently, \textbf{quality metric} is usually expressed as a formula, in which various facts from the testing process or SUT model is used (e.g. number of executed test cases, number of found defects, number of requirements covered by test cases, measured times in the tests and others). Also, elements of SUT models used for test design purposes can be used (e.g. various metrics capturing structural test coverage).

In this paper, we attempt providing a unified classification of quality characteristics specifically dedicated to IoT architecture, ranging from the availability to the cyber--security and usability of the systems.

The remainder of this paper is organised as follows; Section~\ref{RelatedWork} broadly introduces work on quality characteristics and metrics, while in Section~\ref{Classification} we provide a unified classification of quality characteristics focusing on IoT systems. Section~\ref{Discussion} records selected points related to the consolidation process. The last Section concludes the paper.

\section{Related Work}
\label{RelatedWork}

Currently, a set of individual attempts to categorize quality characteristics for IoT systems can be found, most of them focusing on a specific area or aspect of IoT systems, or not sufficiently focusing on the physical level of an IoT system. In this chapter, we discuss these works.

The TMap Suite (previously TMAP Next) is the body of knowledge for professional testers, created by Sogeti corporation, summarized quite an exhausting list of quality characteristics with the selection of these characteristics relevant to IoT testing \cite{TMAP}. However, as the major background of the company is in software testing, this list does not reflect some important networking and physical layer aspects of IoT systems. We can find these characteristics covered by other authors.

A quality model targeting cloud service called Cloudqual was defined by Zheng et al. \cite{zheng2014}. This model is used to represent, measure, and compare the quality of the cloud service providers. It contains six quality dimensions in total --- one subjective, i.e., usability, and the others objective --- availability, reliability, responsiveness, security, and elasticity. Empirical case studies on three storage clouds were conducted. Similarly, the trust of cloud service providers is calculated using the proposed novel trust model based on past credentials and present capabilities of a cloud resource provider by Manuel \cite{manuel2015}. Quality characteristics discussed in these studies related to the IoT cloud services can be used in the classification of these characteristics for general IoT systems.

Data quality metrics in pervasive environments were defined by Li et al. \cite{li2012} and applied on real--world data sources to demonstrate the feasibility of the metrics. Previous data quality characteristics in database applications were not applicable to pervasive environments and the metrics proposed in QoC research were either unobservable or unadaptable to application requirements. Three metrics were redefined for pervasive environments, namely currency, availability and validity, to quantitatively observe the quality of real--time data and data sources.

Regarding the security area, which is being frequently discussed as one of the most significant IoT challenges \cite{kiruthika2015software,bures2018}, individual studies can be found. As an example, a study by Islam and Falcarin \cite{islam2011} can be given.  
The authors identified security requirements through asset--based risk management process to describe the security goal. Security of the IoT platform is one of the most important requirements, and the results of this study are applicable here.

Various security metrics are used to quantify the degree of freedom from a possibility of suffering damage or loss from a malicious attack. These key metrics have been defined by Abbadi \cite{abbadi2011security}. Security and usability assessment of several authentication technologies are analyzed and summarized by Solie \cite{sollie2005security}.

The applications of IoT bring new possibilities of what the user can achieve and experience. A subjective user's satisfaction with the application --- quality of experience, QoE --- will become new quality metrics the operators will have to consider. Ikeda et al. propose a framework of scalable QoE modeling for explosively increasing applications \cite{Ikeda2016}. They defined two sets of metrics --- physical metrics emerging in the IoT architecture and metaphysical metrics demanded by users.

The quality of the data at the device and network level is also covered by the literature. Banerjee and Sheth explore challenges in interpreting and evaluating the quality of data at informational and application levels \cite{Banerjee2017}. Authors propose solutions of at the different OSI layers to understand the factors affecting the quality of the system. 

Cloud applications can scale up on down on demand (elasticity) depending on the application load. This characteristic is discussed in the study by Han et al. \cite{han2014}. In this study, elastic scaling approach making use of cost--aware criteria to detect and analyze the bottlenecks of the cloud--based applications along with adaptive scaling algorithm for cost reduction was presented.

The nature of the IoT platform where devices (especially mobile devices) are dynamically joining or leaving the network creates new issues in enforcing QoS of such platform. Gomes et al. discuss this scalability characteristic and propose a new approach for resource allocation focusing on the performance of the system when participants disconnect \cite{gomes2018}.  

Another relevant characteristic, an information flow efficiency is explored in supply chain management by Badenhorst et al. \cite{badenhorst2013developing}. A conceptual framework of indicators and data--oriented metrics to evaluate the efficiency of information flows in supply chains are introduced in this study.

Also testability of an IoT system, especially testability by automated tests shall be considered as a quality characteristics. Previous attempt to define metrics for automated testability has been done for web applications \cite{bures2015metrics,bures2015model}, relevant for automated tests using the web--based user interface of the SUT. As IoT systems provide web-based user interfaces in many cases, this proposal can be applicable also to IoT context. 

As the individual works discussed in this section focus rather on the IoT specific areas, on certain segments of the whole IoT platform, or does not reflect the quality aspects of IoT system in their full spectrum, a consolidated view has to be created to cover the whole spectrum of the IoT quality characteristics.

\section{Proposed Classification}
\label{Classification}

In the proposed classification, we merged several relevant proposals \cite{zheng2014,li2012,TMAP,islam2011,Ikeda2016,Banerjee2017,manuel2015,han2014,buyya2009,gomes2018,abbadi2011security,badenhorst2013developing,sollie2005security} into one unified view, which we enriched by several own suggestions of quality characteristics relevant for the IoT systems. 

In the proposal, we followed several design principles:

(I) We added a physical device layer aspect to the classification, as this aspect becomes especially relevant in case of the IoT systems.

(II) We focused in special detail on Security, Privacy and Usability aspects, as these areas are considered as being critical for the IoT domain \cite{kiruthika2015software,bures2018}.

(III) We tried to minimize possible overlaps and duplications in the final proposed classification.

In this paper, we deliberately focus on quality characteristics instead of more detailed quality metrics. The reasons are the following: (1) the quality metrics might be too specific considering particular subdomain of IoT systems so that generalization might be not possible, and (2) much more SUT modelling information shall be required, making such attempt being out of the scope of the conference paper. Hence, in our consolidation, we abstracted some of the quality metrics from a subset of surveyed work (for instance \cite{abbadi2011security,Ikeda2016}) to the quality characteristic without biasing the original purpose and meaning of the metric.

Table \ref{tab.quality_characteristics} presents this consolidated view. Regarding the level of granularity, we decided to use two levels: main level quality characteristic (in Table \ref{tab.quality_characteristics} by \textbf{bold}) and second--level quality characteristic, being a subcategory of the main level (in Table \ref{tab.quality_characteristics} by \textit{italics, indented}).

For several quality characteristics, synonyms have been used in the investigated literature. Also, due to our experience, several synonyms are used in the industry praxis. We put these synonyms to the footnote with a citation to the source paper (or a comment that the synonym is our suggestion based on the industrial praxis). The last column of the Table \ref{tab.quality_characteristics} presents the origin of the suggested quality characteristic. Word \textit{own} in this column indicates that the quality characteristic is our suggestion based on the industrial experience and quality characteristics defined in test management methodologies for traditional software systems.

%{.56\textwidth}{.56\textwidth}
\begin{longtable}{|l|p{.47\textwidth}|c|}
\caption{Proposed unified classification of IoT systems Quality Characteristics}
\label{tab.quality_characteristics}
\endfirsthead
    \hline 
    \textbf{Quality characteristic} & \textbf{Description} & \textbf{Source} \\
    \endhead
    \hline  \levelone{Accessibility} & The extent to which the system can be handled by users with specific needs. & own  \\
    
    \hline  \levelone{Availability} & Availability of the provided service or particular data supplied as the part of the service. & \cite{zheng2014,li2012} \\ 
    
\hline  \levelone{Device lifespan expectancy} & What is the estimated lifetime of a single HW device? & own \\ 
    
\hline \levelone{Elasticity} & The ability of the system to provide particular service on demand during a time interval. & \cite{zheng2014}  \\ 
    
\hline  \levelone{Installability} & What is the estimated lifetime of a single HW device? & \cite{TMAP}  \\ 
    
\hline  \leveltwo{Ease of deployment} & Effectiveness and efficiency with which the application can be deployed to devices of the system.& own  \\ 
    
\hline  \levelone{Interoperability\footnote{also Compatibility \cite{TMAP}}} & Capability of the system, product, or device to interact with another system, product or device or to interchange data with it.& own, \cite{TMAP} \\ 

\hline  \levelone{Maintainability} & Effort needed to perform various maintenance tasks of the deployed system. & \cite{TMAP}  \\ 

\hline  \leveltwo{Replaceability} & Effectiveness and efficiency with which an invalid unit or device of the system can be replaced & own \\ 

\hline  \leveltwo{Updateability} & Effectiveness and efficiency with which a unit or device be updated to the latest version. & own \\ 

\hline \leveltwo{Performance} & The extent to which the system is able to handle a certain amount of data and concurrent user/device traffic. & own \\ 
    
\hline \levelone{Privacy}  & The extent to which the system maintains access to the user data corresponding to defined access rights by all involved parties and the extent to which the system prevents abuse of the user data & \cite{TMAP,islam2011}  \\   
\hline \leveltwo{Randomness}  &  The extend to which the cryptographic algorithm used for protection generated random numbers (i.e. entropy size, or the randomness of the seed). &\cite{islam2011}\\      

\hline \leveltwo{Data privacy\footnote{also Confidentiality \cite{TMAP} or Data Store \cite{islam2011}}}  & The extend to which the data is safely stored with appropriate measures (i.e. Encyrption). This can also include the location of storage. & \cite{TMAP,islam2011} \\   

\hline \leveltwo{Data Transmission}  & The extend to which the data is vulnerable to a replay or a Man in the Middle (MITM) attack. The extent to which the data being transmitted is encrypted with an appropriate algorithm &\cite{islam2011} \\   
\hline \leveltwo{Access Control}  & The extend to which the user has access to data, and the data he and others can modify.& \cite{islam2011} \\   
\hline \leveltwo{Non--Repudiation}  & The extend to which the system can guarantee that the data has not been modified &\cite{islam2011} \\   
\hline \leveltwo{Proof of Transaction}  &  The extend to which a transaction can be proven to be from a user and the extent to which this user can be identified by other users if the data is leaked. & \cite{islam2011}\\

\hline \levelone{Reliability\footnote{also  Correctness (our suggestion)}}  & The extent, in which the system is free from hardware and software defects, or other defects, which can lead to system failures.& \cite{TMAP,zheng2014} \\   
\hline \leveltwo{Data quality\footnote{also Precision \cite{li2012} or Data Integrity  \cite{Banerjee2017}}}  & Is the quality of the data provided by the system on the various levels (device, network, computing, and user interface) sufficient to enable correct run of the service? & \cite{Ikeda2016,Banerjee2017,li2012} \\   
\hline \leveltwo{Functional Correctness}  & What is the error rate of the system in the sense of functional defects affecting the system processes and the procedures handling the data stored in the system? & own \\   
 
\hline \leveltwo{Up--to--dateness\footnote{also Data validity (our suggestion)}} & Are the data obtained from the system or device actual enough to enable correct operation of the service? & \cite{li2012} \\   
\hline \leveltwo{Trustworthiness}  & The extent to which the data provided by the system are trusted by its' users.& \cite{manuel2015,li2012} \\

\hline \levelone{Resource utilization\footnote{also Efficiency \cite{TMAP}}}  & The extent to which the resources required by the system were used in relation to the accuracy and completeness with which users of the system achieve their goals.& \cite{TMAP} \\   
\hline \leveltwo{Estimated energy efficiency}  & How long can the device operate without a power source? Does the device have a backup power source? Can it switch to passive mode when not needed? & own\\   
\hline \levelone{Responsiveness\footnote{also Time--behavior \cite{TMAP}}}  & The extent to with the system handles a request within a required time interval.& \cite{zheng2014} \\   
\hline \levelone{Satisfaction}  & The extent to which user needs are satisfied when the system is used in a particular operation. &\cite{TMAP} \\

\hline\levelone{Scalability} & The extent to which the system can adapt to new operational conditions, as the deployment model, size of processed data, user traffic conditions, added or removed devices and others.&\cite{han2014,buyya2009,gomes2018} \\   
\hline \levelone{Security} & The extent to which the system protects its data so that any other party accessing to the system is given a level of access to this data, which is appropriate to the particular level of authorization.&\cite{TMAP,zheng2014,abbadi2011security} \\
\hline \leveltwo{Attack Surface}  &  The number of interfaces provided to the user to access data and their associated security.&\cite{abbadi2011security} \\   

\hline \leveltwo{Given Sense of Control}  & The extend to which the user has access to control the device/application and the associated data collected and shared with the third party.&\cite{abbadi2011security} \\   
\hline \leveltwo{Flaws vs Time}  & The number of critical flaws found over a period during a review or after deployment.& \cite{abbadi2011security} \\

\hline \leveltwo{Data Timeliness}  & The extend to which the data are backed up and can be retrieved by the users and/or forensic investigators.& \cite{badenhorst2013developing} \\   
\hline \leveltwo{Data Provenance}  & The extend to which the data is guaranteed to be provided by a trusted source. & own \\
\hline \leveltwo{Security Compliance}  & The extend to which a system is compliant with a given security standard fit for its purpose (i.e. Critical infrastructure, Military, General Public).& \cite{abbadi2011security} \\   

\hline \levelone{Testability}  & The extent how easy is to design and conduct tests for the system, especially automated tests.&own \\

\hline \levelone{Usability}  & The extent how easy, efficient, and enjoyable the interface of the system is to use and how efficiently the user interface contributes to support of the tasks user has to perform in the system.&\cite{TMAP,zheng2014,sollie2005security} \\   
\hline \leveltwo{Subjective Satisfaction}  & The extend to which a user is satisfied with both the software and its interface&    \cite{sollie2005security} \\   
\hline \leveltwo{Rate of User Error}  & The extend to which the user encounter errors on the system or is required to perform an error action (i.e. reset a password, back--end error).&\cite{sollie2005security} \\   
\hline \leveltwo{Speed to Learn} & The time required by the user to learn about the software and intrinsic characteristics. This can relate to the time for the user to perform an easy, medium or hard action with a software.&\cite{sollie2005security} \\   
\hline \leveltwo{Support of Secure Behavior}  & The extend to which the security notifications and control are both enjoyable and understandable by a lay user. The user interface contributes to the cyber--situational awareness of the user. & own \\   
%\hline   &  \\ 

    \hline 
\end{longtable}

\section{Discussion}
\label{Discussion}

In this section, discuss several issues related to taken approach and related questions.

Regarding the selection of the resources, to compile the presented consolidated view, we preferred works which are also consolidating the previous ideas, for instance, summary by Li et al. \cite{li2012}, which aggregates a set of previous works as \cite{bu2006managing,kim2006quality,sheikh2007middleware,manzoor2008evaluation,neisse2008trustworthiness}.

During the creation of the presented consolidated view on IoT systems Quality Characteristics, interpretation of particular items may be different by individual authors. As an example, \textit{Availability} can be discussed: this characteristic is described as "Uptime percentage of cloud services during a time interval" by Zheng et al. \cite{zheng2014}, or as a "Availability of the data sources, measured by a ratio of the number of attributes available to the total number of attributes" by Li et al. \cite{li2012}. In such cases, we consolidated the metric to more general one, as, in the example of \textit{Availability} was "Availability of the provided service or particular data supplied as the part of the service". Another example of this generalization is the \textit{Data quality}, where Li et al. \cite{li2012} understands this characteristic to cover all layers of the SUT spanning from physical layer to the user interface layer, whereas Ikeda et al. \cite{Ikeda2016} discuss this characteristic in context of the IoT devices. Similarly, we unified a concept of \textit{scalability} understood differently by various authors \cite{han2014,buyya2009,gomes2018}.

In case of \textit{Resource utilization} and \textit{Efficiency} suggested by TMAP \cite{TMAP}, these two categories seem rather overlapping, even if not equivalent. Hence, we decided to merge these both categories into final \textit{Resource utilization}, as this characteristic express the idea better.  

In another situation, we decided to add more specific quality characteristic, than was reported in works dealing with this topic previously. The example is \textit{Accessibility}. In \cite{TMAP} and \cite{zheng2014}, \textit{Accessibility} is implicitly understood as a part of \textit{Usability}, however, according to the common understanding of these two concepts, for instance \cite{Henry:2014}, our suggestion is to distinguish these two categories.

We made also generalization in case of \textit{Confidentiality} suggested by TMAP \cite{TMAP} --- we included this characteristic as a subtype of \textit{Privacy} category, named as \textit{Data Privacy}, as we merged this characteristic with \textit{Data Store} suggested by Islam et al. \cite{islam2011}. 

A discussion can be made, if suggested \textit{Up-time} subcategory of the \textit{Reliability} does not duplicate \textit{Availability}. As the \textit{Up-time} related to particular IoT devices in the sense of their reliability, whereas \textit{Availability} category describes the overall availability of the system, we decided to keep these two characteristics separated. 

In the proposed categorization we decided to exclude characteristics metrics related to test coverage levels \cite{van2013tmap} as well as metrics for assessment of efficiency of test cases, for example \cite{Bures2017,li2012better}. Such metrics might be discussed in the context of testing process efficiency, however, does not directly relate to the quality characteristics of SUT  (a relevant exception would be, when an automated test suite was considered as an SUT).

Another point can be raised regarding a question, if proposed two--level categories are appropriate to organize the discussed quality characteristics. In the current categorizations, only one level list is usually used, for instance, \cite{jung2004measuring} or \cite{van2013tmap}. However, specific focus on physical layer aspects, security, privacy, reliability and maintainability of an IoT system led us to the identification of more relevant subcategories, which justify the proposed two--level structuring.

\section{Conclusion}
\label{Conclusion}

The usage of quality characteristics contributes to the better structuring of the testing process, helps in the test reporting and acts as a check--list for the test engineers aiding decision which quality aspects of the SUT to test. For these reasons, we consider it useful to provide a comprehensive view of the quality characteristics for the IoT systems, focusing on the specifics of these systems. As the previous work discusses rather individual areas of IoT systems and particular subareas of system quality, in this paper we provide a consolidated view. This effort involved extensive discussions arising from an attempt to consolidate the particular terminology used by various authors; we summarize this discussion in the Section \ref{Discussion}. In the proposed classification we emphasized specific characteristics of the IoT system. Namely, we reflected physical device layer more intensely in comparison to standard software quality characteristics, for instance, \cite{jung2004measuring,van2013tmap} and we focused in special detail to Security, Privacy and Usability aspects, as these areas are considered as being critical for the IoT domain \cite{kiruthika2015software,bures2018}. This focus makes the proposed IoT characteristics more relevant to IoT systems, compared to a case, when standard software quality characteristics would be used in testing of an IoT solution. 

\bigskip

\section*{Acknowledgements}
This research is conducted as a part of the project TACR TH02010296 Quality Assurance System for Internet of Things Technology.

\bibliographystyle{splncs04}
\bibliography{bibliography}

\color{blue}

\bigskip\bigskip\bigskip\bigskip
\bigskip
\large
 
Paper accepted at 
\bigskip 

\textbf{Urb-IoT 2018}

\textbf{3rd EAI International Conference on IoT in Urban Space}

\bigskip November 21-23, 2018, Guimarães, Portugal

\bigskip 

http://urbaniot.org/full-program/

\color{black}
\normalsize

\end{document}